\documentclass[a4paper]{jpconf}
\usepackage{graphicx}
\graphicspath{{figures/}}
\begin{document}
\title{Micro-physics simulations of columnar recombination along nuclear recoil tracks in high-pressure Xe gas for directional dark matter searches}

\author{Y~Nakajima,  A~Goldschmidt, M~Long, D~Nygren, C~Oliveira, J~Renner}

\address{Lawrence Berkeley National Laboratory, Berkeley, CA 94720, USA}

\ead{YNakajima@lbl.gov}


\begin{abstract}
Directional sensitivity is one of the most important aspects of
WIMP dark matter searches. Yet, making the direction of nuclear
recoil visible with large target masses is a challenge. To
achieve this, we are exploring a new method of detecting directions of
short nuclear recoil tracks in high-pressure Xe gas, down to a few
micron long, by utilizing columnar recombination. Columnar
recombination changes the scintillation and
ionization yields depending on the angle between a track and the
electric field direction. In order to realize this, efficient cooling of
electrons is essential. Trimethylamine(TMA) is one of the candidate
additives to gaseous Xe in order to enhance the effect, not only by efficiently cooling the
electrons, but also by increasing the amount of columnar
recombination by Penning transfer.
We performed a detailed simulation of ionization electrons transport
created by
nuclear recoils in a Xe + TMA gas mixture, and evaluated the size of the columnar recombination
signal. The results show that the directionality signal can be
obtained for a track longer than a few $\mu$m in some ideal cases. 
Although  more studies with realistic assumptions are still needed in
order to assess feasibility of this technique, this potentially opens
a new possibility for dark matter searches.
\end{abstract}

\section{Introduction}

Discovering the nature of dark matter is one of the most important goals in Particle Physics.
If the dark matter consists of Weakly Interacting Massive Particles
(WIMPs), they would occasionally interact with matter on Earth and make 
low-energy nuclear recoils at O(10)~keV.
The signature of such nuclear recoils would be a small energy deposition
with a preferred direction due to the Solar System's rotation in our
galaxy.
Most of the experiments with leading sensitivity use liquid noble gas
detectors for their scalability and effectiveness of background
rejection. In such detectors, however, searches solely rely on the detection of
a small energy deposition, and detecting the directional signal is
extremely difficult since the track length of a nuclear recoil in
liquid is
expected to be only O(10)~-~O(100)~nm.

On the other hand, there are many ongoing efforts aimed at WIMP dark
searches with directional sensitivity using low-pressure
($\sim$100~torr) gaseous
TPCs. In such detectors, the track length would be O(mm) and can be 
detected with imaging devices. A drawback of this kind of
detectors is that the density of target gas is typically 
$10^{-4}$ times smaller when compared with liquid noble gas detectors,
which makes it very hard to achieve competitive sensitivity.

To fill the gap, we are exploring a new method of detecting
the direction of nuclear recoils using columnar recombination~\cite{Nygren:2013nda}, 
which would be capable of identifying the direction of 
a track several $\mu$m in length in
a high-pressure ($\sim$10~bar) gaseous TPC. The density of target
material can be about 100 times larger than that of the low-pressure gaseous
TPCs designed for directional dark matter searches. 
If this technique is realized, it would be the first detector that
is capable of detecting the directional signal of nuclear recoil
with a target mass large enough to compete with liquid-based TPCs.
In order to test the feasibility of this idea, we performed a
detailed simulation of transport and recombination of ionization
electrons in a high-pressure
gaseous medium.

\section{Directional sensitivity with columnar recombination}

We are exploring directional dark matter searches with high-pressure
gaseous Xenon (Xe) TPCs, and the following study is based on this kind
of TPC. Xenon is an attractive target medium,
because of its high mass and therefore greater predicted spin-independent WIMP scattering cross section,
and also for
the possibility of searching for neutrinoless double beta decay of
${}^{136}$Xe with the same detector.
On the other hand, a similar principle of directional dark matter
searches should be applicable 
to TPCs based on other noble gases, such as Argon or Neon.

\subsection{Columnar recombination}

\begin{figure}[h]
\begin{center}
\includegraphics[width=0.8\columnwidth]{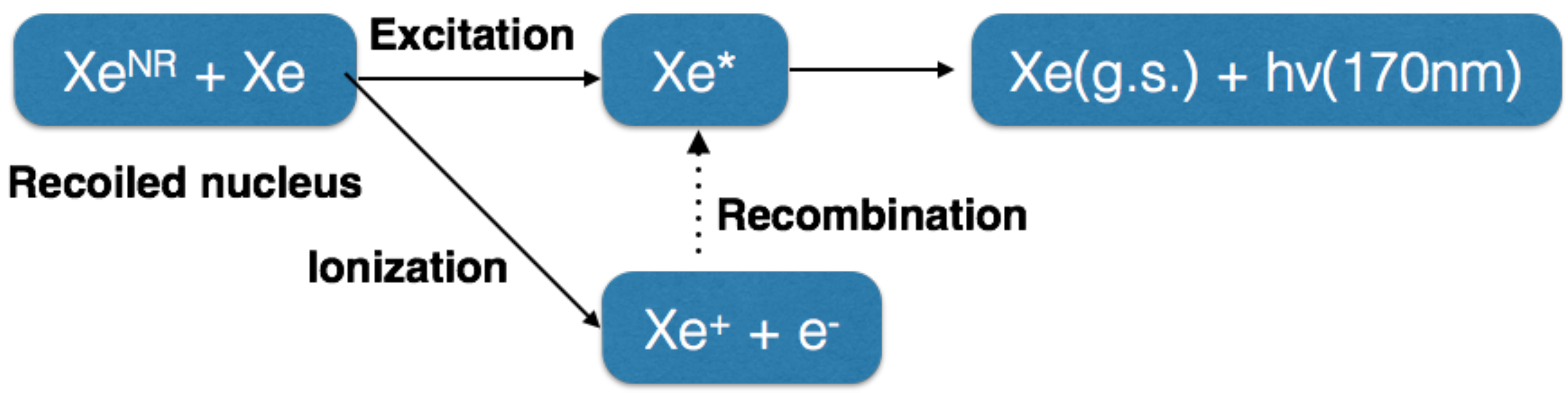}
\end{center}
\caption{\label{fig:xe-reaction}Simplified schematic of interaction of
recoiled Xe nucleus in Xe medium}
\end{figure}

If a WIMP particle collides a Xe atom in the detector medium,
the recoiled nucleus will excite and
ionize other Xe atoms along its track. Figure~\ref{fig:xe-reaction} shows a simplified
schematic of this process. Typical observable signals are
VUV scintillation light from excited Xe and charge from ionized electrons.
The recombination of Xe$^+$ and e$^-$ produces scintillation light as
well, and therefore changes the ratio of
scintillation light and charge yields. 

\begin{figure}[h]
\begin{center}
\includegraphics[width=0.7\columnwidth]{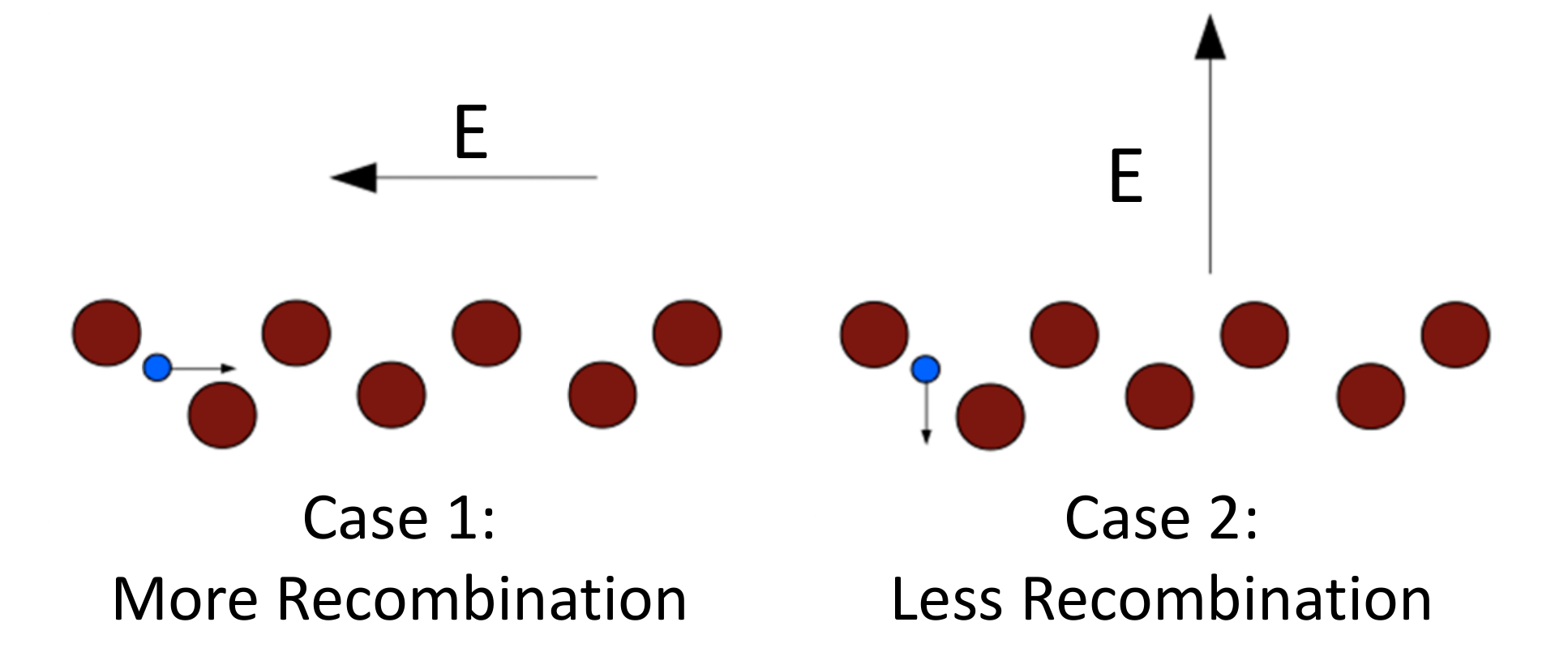}
\end{center}
\caption{\label{fig:columnar-recombination} Schematic of columnar
  recombination. The large brown circles represent ions and the small
  blue circles represent electrons.}
\end{figure}

Columnar recombination is the process by which the ionized electrons
recombine with other non-parent Xe ions along the track. Its
probability depends on the relative angle between the
track and the external electric field. If the track is parallel(perpendicular) to the
electric field, ionized electrons have more(less) chance to meet Xe ions and
therefore recombination would be increased(decreased),
as illustrated in Fig.~\ref{fig:columnar-recombination}.
This is an observed phenomena for
$\alpha$-particles~\cite{TIPP-diana} in a Xe~+~TMA gas mixture.
The goal of this study is to see if 
the phenomena can be observed for
much shorter tracks of recoiled O(10)~keV Xe nucleus.


In order to obtain the directional sensitivity with columnar recombination, the
distributions of both ions and
electrons from the ionization process  need to preserve the directional information 
while they overlap each other. 
We describe these requirements in detail below.

\subsection{Requirement for ions}

\begin{figure}[h]
\includegraphics[width=24pc]{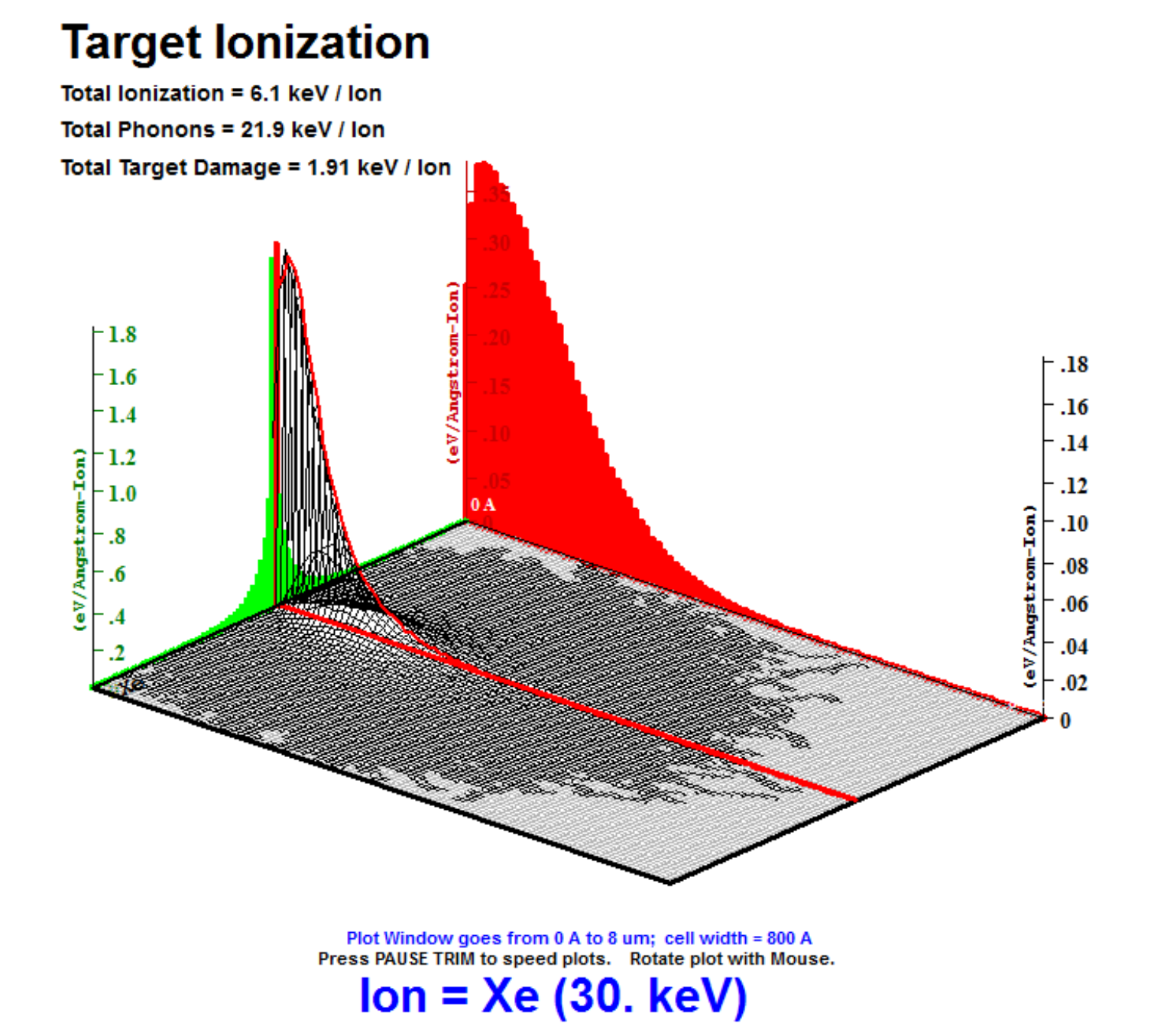}\hspace{0pc}%
\begin{minipage}[b]{14pc}\caption{\label{fig:SRIM}
Simulated distribution of ionization made by 10000 Xe ions of 30~keV,
injected along the X-axis (the red line), shown on a 8$\times$8 $\mu$m plane.
The red(green) histogram shows the distribution of ionization
projected onto the axis parallel(perpendicular) to the original ion
direction.
The Xenon gas density is set to $\rho = 0.05$~g/cm${}^3$, which
corresponds to about 10~bar at room temperature.
This figure was taken from Ref.~\cite{Nygren:2013nda}.
}
\end{minipage}
\end{figure}

Figure~\ref{fig:SRIM} shows a simulated distribution of ionized Xe made by 30~keV
recoiled Xe nuclei in 10~bar Xe gas, calculated using the Stopping and Range
of Ions in Matter (SRIM) simulation
package~\cite{SRIM}.
As shown in this figure, distribution along the original recoil
direction is larger than its transverse direction.
The energy weighed range of 30~keV Xe ions in 10~bar high-pressure Xe
gas is expected to be about 2~$\mu$m.

Ionization electrons are attracted to ions by their Coulomb potential
before they recombine. Therefore, the shape of the Coulomb potential
that ions create is
one of the factors that determine the directional sensitivity.
To quantify this, we evaluate the Onsager radius, $r_0$, which is the
distance
at which the Coulomb potential of the electron-ion system is equal to
the thermal energy $k_BT$,
 where $k_B$ is the Boltzmann constant, and $T$ is the temperature.
It can be simply calculated as $r_0 = e^2/(\epsilon k_B T)$,
where $e$ is the charge of electron, $\epsilon$ is the dielectric
constant.
This $r_0$ roughly corresponds to the resolution of
detecting a distribution of ions by ionization electrons. 

Table~\ref{onsager-radius} shows the typical track length of nuclear
recoils and the Onsager radius for liquid and gaseous Xe. 
In the case of liquid Xe, the expected track length would be roughly
equal to the Onsager radius, and therefore no significant
directionality signal would be detected.
On the other hand, in gaseous Xe, the track length would be much longer than the
Onsager radius, and it should be visible as a track for
ionization electrons. Therefore, it is essential to use gaseous phase
Xe in order to utilize the columnar recombination.
The density of a gaseous phase detector is lower compared to a liquid
phase detector,
but the expected detector size that would be necessary
to compete with currently leading experiments is still within a realistic
scale. For example, the physical size of a 
1-ton target mass with a 10~bar
high-pressure gaseous Xe detector would be about 20~m${}^3$.

\begin{center}
\begin{table}[h]
\caption{\label{onsager-radius}Comparison of typical track length due
  to Xe nuclear recoils of 30~keV kinetic energy and
  the Onsager radius for liquid and gaseous Xe phases.}
\centering
\begin{tabular}{@{}l*{3}{c}}
\br
&Density & Track length & Onsager radius\\
& & (30~keV NR) & $(r_0)$\\
\mr
Liquid Xe& 3.1 g/cm${}^3$ & $\sim 50$~nm & $\sim 50$~nm \\
Gas Xe (10~bar, 300K) &0.05 g/cm${}^3$ & $\sim 2~\mu\textrm{m}$ & $\sim 70$~nm \\
\br
\end{tabular}
\end{table}
\end{center}

\subsection{Requirement for electrons}

The requirements for electrons are more stringent, 
as the directional
information is further washed out due to the diffusion process. 
In addition, electrons need to be thermalized in order to 
efficiently recombine with ions.

Pure Xe would not satisfy this requirement, because it lacks inelastic
scattering processes below its first excitation energy of $\sim$7 eV, and it
will take an extremely long time to thermalize electrons and cause large
diffusion.
Therefore, molecular additives with large inelastic cross sections at
low energy  are
needed. We chose Trimethlamine (TMA) as a candidate additive, 
because it has a large inelastic cross section due to many 
vibrational and rotational 
modes, which would efficiently cool down electrons. In addition,
it is expected to TMA enhances the ionization signal through the Penning effect
($\textrm{Xe}^* + \textrm{TMA} \to \textrm{Xe} + \textrm{TMA}^+ + e^-
$), which contributes to increased columnar recombination. 
We will examine more details about the effect of molecular additives
in the following section.

\section{Micro-physics simulation of recombination}
\subsection{Simulation setup}

In order to understand the electron diffusion process with such molecular additives and
to make a realistic estimation of the recombination signal, we performed a
micro-physics simulation of electron transport in the gas mixture of
Xe and TMA under a uniform external electric field.
The simulation is based on
Garfield++~\cite{Garfieldpp} 
and Magboltz~\cite{Magboltz}
with a custom implementation of the recombination process which takes into account the influence
of the Coulomb field created by ions and electrons themselves. 
The electron interaction cross sections with Xe and TMA in
Magboltz version 9.01  are used for this work.
Only electrons produced via primary ionization are tracked in the simulation, and secondary ionization is not considered.
The distribution of initial electron energy, $dN/dE$, is set to 
$dN/dE \propto 1/(E^2+(7.6\textrm{~eV})^2)$ with a cut-off at 
7~eV, based on Ref.~\cite{Opal:1972}.
The temperature was set to room temperature (293~K). 
An electron was considered to have recombined with an ion if it passed within one de Broglie wavelength of it.
We simulated up to 100 electron and ion pairs simultaneously. 
Simulating more than 100 electron and ion pairs was possible but not
practical because of its computational intensiveness.

\subsection{Size of  electron diffusion}
In order to extract directionality information from columnar
recombination, the diffusion of ionization electrons must be
small enough that their spread is 
smaller than the track length 
while they overlap with the ions.
Assuming a typical drift
velocity of $\sim$O(1) $\mu$m/nsec and a track length of a few $\mu$m, 
the duration of the overlap is expected to be a few nsec after the initial ionization.

\begin{figure}[h]
\begin{center}
\includegraphics[width=18pc]{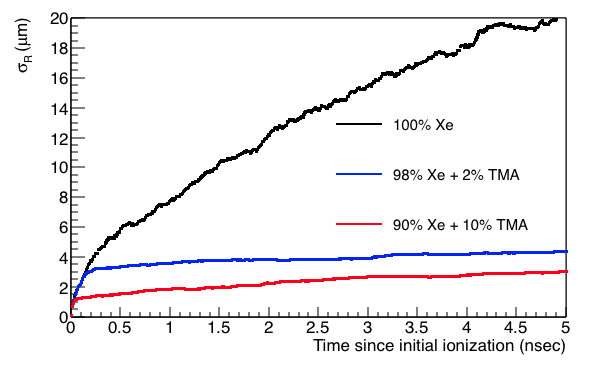}
\includegraphics[width=18pc]{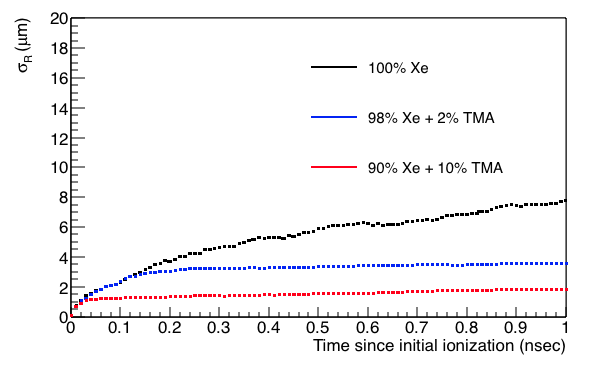}
\caption{\label{fig:electron-diffusion} 
  The simulated size of $\sigma_R$  as a function of time since
  initial ionization 
  for various Xe + TMA mixtures at a
  total pressure of 10~bar. Results from pure Xe (black), a 98\% Xe and
  2\% TMA mixture (blue) and a 90\% Xe and 10\% TMA mixture (red) are
  shown.
  The left and right panels show the same data over different horizontal axis ranges.
}
\end{center}
\end{figure}

Figure~\ref{fig:electron-diffusion}  shows the size of 
the standard
deviation of simulated electron location perpendicular to the electric field,
$\sigma_R$, as a
function of the time since the initial ionization at 10~bar. The results
show that the size of diffusion
can be significantly reduced with the addition of TMA. With the gas
mixture of 10\% TMA and 90\% Xe, the size of $\sigma_R$ can be
reduced to be $\sim$2~$\mu$m for 5~nsec after the initial ionization.

As shown in the right panel of Fig.~\ref{fig:electron-diffusion}, the
diffusion process can be  separated into two distinct phases; the
initial $\sim$0.1 nsec with quick expansion where the effect of TMA is
marginal, and the following relatively stable phase. 
The time to reach the stable phase tends to be shorter with increased
TMA concentration. 

\begin{figure}[h]
\begin{center}
\includegraphics[width=18pc]{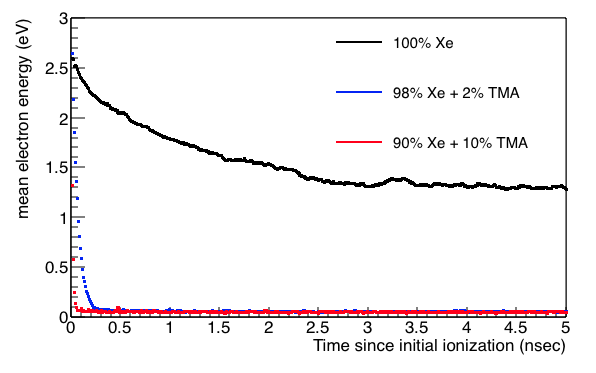}
\includegraphics[width=18pc]{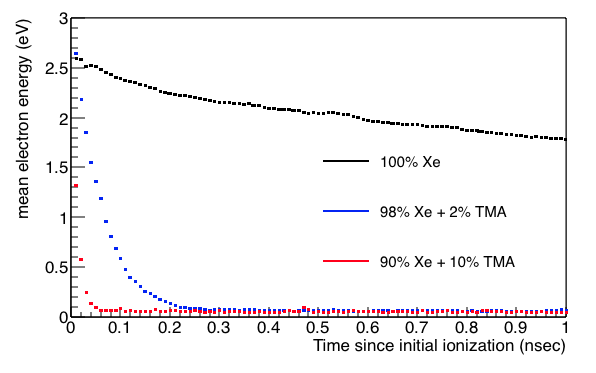}
\caption{\label{fig:electron-ke}
  The simulated size of the average kinetic energy of ionization
  electrons as a function of time since initial ionization  for various Xe + TMA mixtures at the
  total pressure of 10~bar. Results from pure Xe (black), a 98\% Xe and
  2\% TMA mixture (blue) and a 90\% Xe and 10\% TMA mixture (red) are
  shown.
  The left and right panels show the same data over different horizontal axis ranges.
}
\end{center}
\end{figure}

This phase transition is due the thermalization  process of the
electrons. Figure~\ref{fig:electron-ke} shows the average electron kinetic
energy as a function of the time since the  ionization. TMA is expected to cool
electrons quite efficiently, and the average kinetic energy of
electrons would
reach the thermal energy of $\sim$0.025~eV within $\sim$0.1 nsec.
Therefore, adding TMA to
the gas mixture would shorten the initial expansion phase before
thermalization and would help to keep ionization electrons 
close to the ions.

\subsection{Expected recombination signal}

The result in the previous section shows that the electrons are
expected to be kept within $\sim$4(2)~$\mu$m of the track for a 2(10)\%
mixture of TMA at 10~bar total pressure.
This suggests that directional sensitivity from columnar
recombination can be obtained for tracks longer 
than $\sim$4(2)~$\mu$m with a TMA concentration of 2(10)\%.
The actual recombination probability would be a complex function of
both TMA fraction and the strength of the external electric field. 
To estimate this function, we simulated the transport of 100 electron-ion pairs under the
following initial conditions:
\begin{itemize}
\item A 1 $\mu$m track with 100 equally separated electron-ion pairs
  aligned either
  parallel or perpendicular to the external electric field. (10~nm 
  spacing)
\item A 4 $\mu$m track with 100 equally separated electron-ion pairs
  aligned either
  parallel or perpendicular to the external electric field. (40~nm 
  spacing)
\end{itemize}

For the first set of simulations with a 1 $\mu$m track length, 
the 10~nm spacing was chosen since 
  it is approximately the same as that of $\alpha$ particles and
  is a realistic ionization density for a nuclear
  recoil, 
even though the track
length is shorter than the $\sigma_R$ and no directionality signal is
expected.
On the other hand, while the ionization density is less realistic, we
expect some directionality signal in the simulations of 4~$\mu$m
tracks since the track length would be equal to or longer than $\sigma_R$ for the
simulated Xe + TMA mixtures.

\begin{figure}[h]
\begin{center}
\includegraphics[width=0.8\columnwidth]{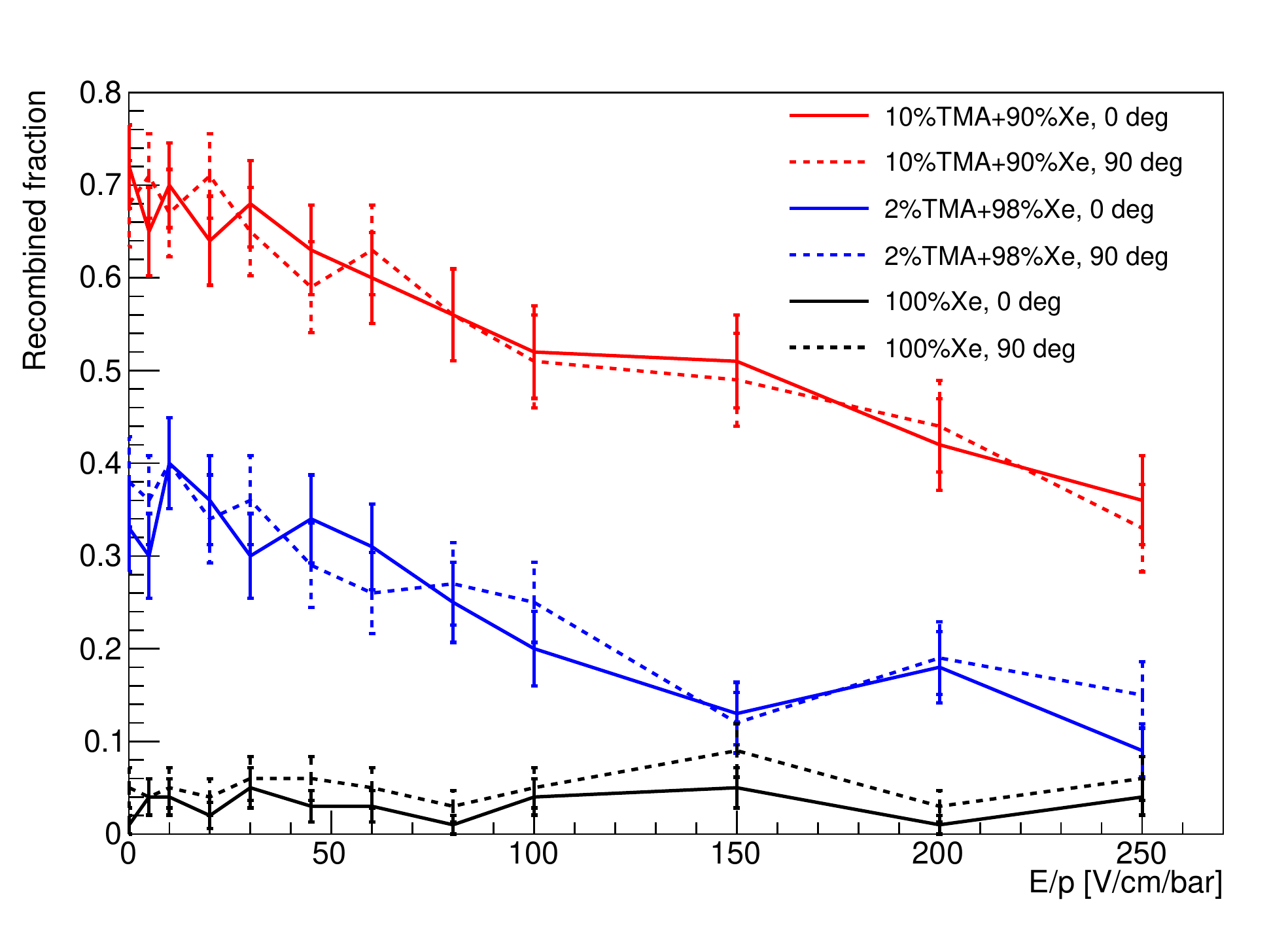}
\includegraphics[width=0.8\columnwidth]{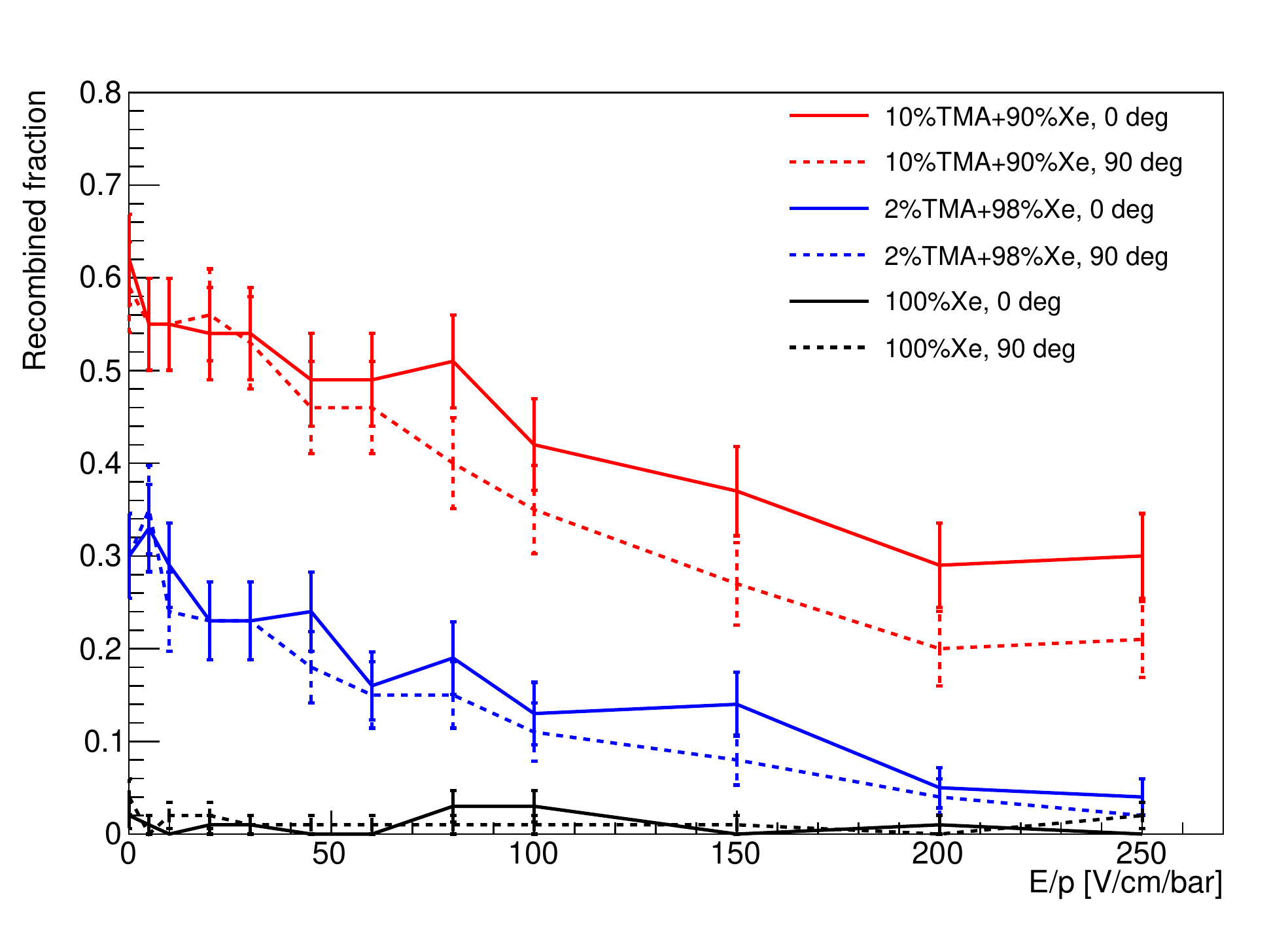}
\end{center}
\caption{\label{fig:recomb-frac}
  The simulated size of recombination fraction, (total number of
  recombined electrons) / (total number of initial ionization electrons), from 100 
  electron-ion pairs in a 1~$\mu$m track (top panel) and a 4~$\mu$m track (bottom panel).
  The results for various  reduced external electric field and various Xe + TMA mixtures at the
  total pressure of 10~bar are shown. 
  The solid(dashed) lines shows the results with the initial track parallel(perpendicular) to the external electric fields.
  The black, blue, and red points are results with pure Xe, 
  a 98\% Xe and 2\% TMA mixture,  and a 90\% Xe and 10\% TMA mixture, respectively.
  The error bars show the statistical uncertainty of a single event assuming that
 both total number of recombination photons and initial ionization
 electrons are measured with  100\% efficiency.
}
\end{figure}

Figure~\ref{fig:recomb-frac} shows
the simulated recombination fraction from the 1~$\mu$m and 4~$\mu$m tracks in  various Xe
+ TMA mixtures and electric fields.
The simulation predicts significant enhancement of recombination 
with the addition of 
TMA for both 1~$\mu$m and 4~$\mu$m tracks. This recombination
fraction naturally decreases as the external electric field gets
stronger. 
As expected, we see no sign of directional sensitivity for the
1~$\mu$m tracks in any combination of gas mixtures and external fields.
On the other hand, we found that the simulation predicts a statistically
significant difference of the recombination fraction between 0-degree (parallel to the
external field) and 90-degree (perpendicular to the external field)
cases
for the 4~$\mu$m tracks in the 10\% TMA and 90\% Xe gas mixture at $E/p >
100$~(V/cm/bar). We also found some directional sensitivity for 
the  2\% TMA and 98\% Xe gas mixture at $50 < E/p < 150$~(V/cm/bar). Those
 are consistent with the expectation based on the size of $\sigma_R$
 described in the previous section. 

Although more realistic assumptions and benchmarks with the
experimental data are needed, these simulation results show that it may be possible to detect directionality of short tracks
of a few $\mu$m by utilizing columnar recombination.

\section{Discussions}

Although the results from the previous section are encouraging, 
many other things must still be considered to realistically estimate
sensitivity to directional information. 

First, this study assumed that the initial ionization was aligned on a
straight line. However, the actual distribution is expected to be smeared as
shown in Fig.~\ref{fig:SRIM}, which will reduce  the directional
information.

Another idealization made was the assumption of 100\% detection efficiency 
for both scintillation light and ionization electrons.
While detecting ionization
electrons at high efficiency is relatively easy, the detection
efficiency of primary scintillation light tends to be small and it is $\leq$O(10)\% in
actual experimental apparatus,
because of a typically limited solid angle coverage and the quantum efficiency of
photon detectors. 
Therefore, the uncertainty of the results shown in
Fig.~\ref{fig:recomb-frac} would be
larger in any realistic detector configuration.

The energy distribution of ionization electrons is also poorly understood. 
There are several  previous  measurements on 
ionization electron energy spectra from 
Xe~\cite{Opal:1972,Toburen:1974vn,Chaudhry:1989jk},
but there are no data points below a few eV, which is the most
important energy region for columnar recombination.
In addition, Penning transfer, if exists, adds a lower energy component to the
electron spectrum.
This would make a significant impact on the 
size of the initial expansion of the electron spatial distribution, and
hence the size of columnar recombination. 
Additional inputs from experimental data are extremely important.

Finally, it should also be noted that recent experimental studies revealed that TMA 
significantly reduces scintillation light yield when it is mixed with
Xe gas~\cite{Yasu-teapot}. Therefore, it is likely difficult to use the gaseous mixture
of Xe and TMA for dark matter
searches. On the other hand, any molecular additives that have a large
inelastic cross section that is similar to TMA would enhance columnar
recombination as well. 
Therefore, it is essential to look for other candidate molecular
additives.

Those factors must be taken into account in estimating realistic
directional sensitivity. 
We expect that there will be still many technical
challenges in order to realize this idea.

\section{Summary}

We are exploring a novel idea of detecting the direction of a very short track in
high-pressure Xe gas by utilizing columnar recombination. 
We performed a micro-physics simulation of ionization electrons to
study this possibility. The results show that,
with sufficient
molecular additives such as TMA which efficiently cool electrons,
 it may be possible to
detect the directionality of a few $\mu$m track in a high-pressure Xe gas.
While more realistic assumptions need to be made in order to evaluate
the feasibility of this technique, it is an encouraging start and
opens a new possibility for WIMP dark matter searches.

\section{Acknowledgments}
This work was supported by the Director, Office of Science, Office of Basic Energy Sciences, of the US Department of Energy under contract no. DE-AC02-05CH11231. J. Renner (LBNL) acknowledges the support of a US DOE NNSA Stewardship Science Graduate Fellowship under contract no. DE-FC52-08NA28752.

\section*{References}

\bibliographystyle{iopart-num}
\bibliography{../TeaPot/nakajima_TPC2014}

\end{document}